\documentclass[11pt,a4paper,english]{amsart}

\usepackage{url}

\usepackage[english]{babel}
\usepackage[]{natbib}
\usepackage{geometry}
\usepackage{graphicx}
\usepackage{caption}
\usepackage{subcaption}
\usepackage{wrapfig}
\usepackage{amsmath}	
\usepackage{amssymb}	
\DeclareGraphicsExtensions{.eps,.png,.pdf}

\theoremstyle{definition}

\theoremstyle{remark}

\numberwithin{equation}{section}

\begin{document}

\title{LIGO-Virgo events localization as a test of gravitational wave polarization state}

\author{Fesik L. E.}
\address{Saint-Petersburg State University, Saint-Petersburg, Russia}
\email{lucia555@yandex.ru}

\author{Baryshev Yu. V.}
\address{Saint-Petersburg State University, Saint-Petersburg, Russia}
\email{yubaryshev@mail.ru}

\author{Sokolov V. V.}
\address{Special Astrophysical Observatory of RAS, Nizhnij Arkhyz, Russia}

\author{Paturel G.}
\address{CRAL-Observatory de Lyon, Saint-Genis Laval, France}


\date{June 14, 2017}


\begin{abstract}
The detection of the gravitational wave events GW150914, GW151226, LVT 151012 and GW170104 by the Advanced LIGO antennas has opened a new possibility for the study of fundamental physics of gravitational interaction. We suggest a new method for determining the polarization state of a
gravitational wave, which is independent of the nature of a GW source. For this, we calculate the allowed sky positions of GW sources along apparent circles. This is done for each polarization state by considering the sensitivity pattern of each antenna and relative amplitudes of detected signals. The positions of circles are calculated with respect to the line joining both LIGO antennas using the observed arrival time delay of the signal between them. The apparent circles (AC) on the sky for allowed positions of the GW sources for the GW150914, GW151226 and LVT151012 events are parallel to the plane of the disc-like large scale structure known as the Local Super-Cluster (LSC) of galaxies which extends up to radius $\sim 100$ Mpc and having thickness $\sim 30$ Mpc. 
For the GW170104 event, the AC is perpendicular to the LSC plane but the predicted position of the source may also belong to the LSC plane, which is consistent with detection of possible optical counterpart ATLAS17aeu.
The next aLIGO-aVirgo observing runs are proposed to test the possibility of clustering the GW sources along the LSC plane.
\end{abstract}

\maketitle

\section{Introduction}
\label{intro}

Sixty years ago, at the 1957 Chapel Hill conference, Richard Feynman argued that a gravitational wave (GW) antenna could in principle be designed that would absorb the energy carried by the gravitational wave (\citealt{feynman95}, p.XXV). Thirty years later, in the 1980's the Laser Interferometer gravitational wave Observatory (LIGO) was proposed for detecting gravitational waves (\cite{LIGO}) with the principal goal to study astrophysical GWs and stimulate research in fundamental physics concerning the nature of gravity (\citealt{abram92}). Recent detection of gravitational wave signals by Advanced LIGO antennas (\citealt{abbott16a}, \citealt{abbott17})
has opened such possibility for study physics of the gravitational interaction. 

In the situation when there is no reliable optical (and other electromagnetic bands) identification of the GW events, the interpretation of the physics of the GW source is still uncertain. Even though the model of tens solar masses binary black holes coalescence at the distance $400 \div 1000$ Mpc is consistent with existing GW data (\citealt{abbott16a}, \citealt{abbott16b}), one should also test alternative possibilities which are allowed by modern theories of the gravitational interaction (\citealt{eardley73}, \citealt{maggiore2000}, \citealt{yunes13}, \citealt{gair13}, \citealt{will14}, \citealt{baryshev17}).

Here we develop a method based on very general physical arguments, which allows us to distinguish between tensor and scalar polarization states predicted by the scalar-tensor gravitation theories. In particular, for the case of two antennas
actual position of the GW source at the apparent circle on the sky can be used for determination of the polarization state 
of the detected GW. Hence, our method allows to test
astrophysical models proposed for explanation
of the physical processes generating the gravitational waves. 

\section{Possible polarization states of a gravitational wave} 
\begin{wrapfigure}{l}{.45\textwidth}
    \centering
    \includegraphics[width=0.7\linewidth]{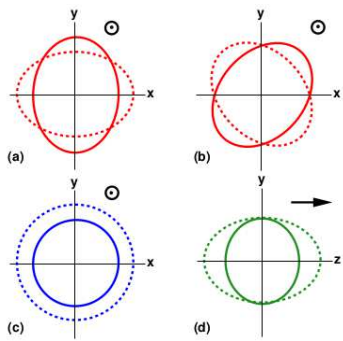}
    \caption{The four (from possible six) polarization modes of GW. There is shown the displacement that each mode induces on a ring of freely moving test bodies. The wave is propagating in the +z direction: in (a), (b), and (c), -- out of the drawing plane; in (d) -- in the plane (\citealt{will14}, \citealt{eardley73}).}
    \label{fig:pstates}
\end{wrapfigure}

In modern theoretical physics, there are two main directions in the study of gravitational interaction: the geometry of Riemannian space and material field in the Minkowski space.

The first approach is the geometrical Einstein's general relativity theory (GRT, also called "geometrodynamics"), which is based on the concept of metric tensor $g^{ik}$ of curved Riemannian space-time (\citealt{einstein15}, the standard textbooks on GRT: \citealt{landau71}, \citealt{misner73}).

The second approach, the Feynman's field gravitation theory (FGT, also called "gravidynamics") is a non-metric relativistic quantum theory, where the gravitational interaction is described by a symmetric second rank tensor potential $\psi^{ik}$ in Minkowski space-time (\citealt{feynman71}, \citealt{feynman95}). The fundamental role of the scalar part of such a tensor, the trace $\psi (\vec{r}, t) = \eta_{ik} \psi^{ik}$, has been demonstrated by \citealt{sokolov80} (see review in \citealt{baryshev17}). 

The classical relativistic gravity effects have the same values in both approaches, however there are also essential differences in observable relativistic astrophysical phenomena 
(\citealt{sokolov15},
\citealt{sokolov93},
\citealt{baryshev95},
\citealt{baryshev17}).

Generally, in modern metric theories based on the metric tensor $g^{ik}$ describing the gravitational potentials, there are six polarization states (\citealt{eardley73}). Three of them are transverse to the direction of propagation, with two representing quadrupolar deformations -- tensor transverse wave, and one representing a monopolar "breathing" deformation -- scalar transverse wave. Other three modes are longitudinal, including a stretching mode in the propagation direction -- scalar longitudinal wave. 

In the frame of GTR, there are only two tensor polarization modes under consideration, ''plus'' ($+$) and ''cross'' ($\times$), whereas some modern modifications of GRT, scalar-tensor gravity theories, predict the existence of separate scalar waves to be detected by means of interferometric (as well as bar) GW antennas.

Concerning FGT, the symmetric second rank tensor potential $\psi^{ik}$ can be decomposed under the Lorentz group
transformations into the direct sum of subspaces: one spin-2, which has five components, one spin-1, with three components, and two spin-0 representations. This decomposition and the appropriate projection operators are exhibited explicitly in \citealt{barnes65}. After taking into account the conservation of the energy-momentum tensor of a GW source, the field theory predicts the existence of three polarization modes: the spin-2
"+" and "$\times$"
tensor transverse waves, and the spin-0 scalar longitudinal waves, which means that FGT is a scalar-tensor gravitation theory. 

It is important to note the principal difference between FGT and scalar-tensor modifications of GRT such as Brans-Dicke theory (hereafter BDT). In the frame of BDT, one introduces an external scalar field $\phi_\textrm{BD}$ having a coupling constant $\omega$, while in FGT the scalar field is the natural internal part of the symmetric tensor field $\psi^{ik}$, i.e. its trace $\psi$, with the Newtonian gravitational coupling constant $G$.


To sum up, according to the mentioned above modern theories of gravitation, there may exist both tensor and scalar polarizations of GW, Fig. (\ref{fig:pstates}), which can be generated by such astrophysical processes as binary coalescence and  collapse or pulsations of massive supernova core. An important task of GW physics is to distinguish between tensor and scalar modes, including their longitudinal and transverse character. In this article, we will show how to use the LIGO-Virgo observations in order to recognise the possible polarization states, and consequently, to get new constraints on fundamental physics of the gravitational interaction.


\section{Determination of GW polarization state}

\begin{wrapfigure}{l}{.48\textwidth}
\centering
\includegraphics[width=.7\linewidth]{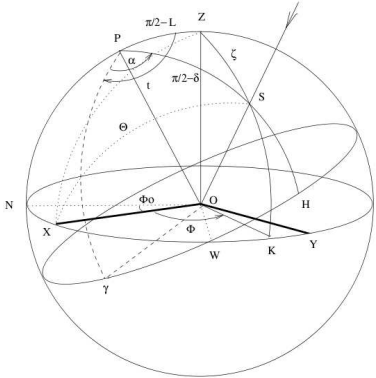}
\caption{Equatorial and horizontal coordinate systems of an interferometric antenna for the GW source S. $Z$ is the zenith, $P$ -- the northern pole,
$\gamma$ defines the sidereal time, $\alpha$ -- the right ascension (RA),
$\delta$ -- the declination (DEC), $\zeta$ -- the zenith angle. The reference direction of the detector is the
direction OX with the azimuth $\Phi_0$.}
\label{fig:sheme}
\end{wrapfigure}

We consider pure geometrical calculations of the detected signal amplitudes for GW having tensor and scalar polarization states. These arguments are very general and strictly determined by the shape of an antenna-pattern, position of the antennas network on the Earth, and the sky-position of the source relative to the corresponding antenna-patterns at a fixed sidereal time of the transient GW event. So considered method does not
depend directly on a particular physics of the GW sources.

\subsection{Method of calculation}

Let us consider a GW antenna based on an Michelson-type interferometer with two orthogonal arms having four test masses at their ends. The receiver is at the rest in the local proper reference frame, with the origin of spatial coordinates in the corner of the system and the X- and Y-axes along the antenna's two arms, Fig. (\ref{fig:sheme}). The GW passing through the antenna displaces the test masses, thereby
 changing the length of each arm from its initial length $L_0$ (for LIGO detectors $L_0 = 4$ km). The monitored by laser difference between lengths of these arms $\Delta L(t) = L_X - L_Y$ gives the observed at the antenna strain $h(t)$:
\begin{equation} \label{eq:1}
    h(t;\zeta, \Phi, \Psi) = \frac{\Delta L(t)}{L_0} = h^0 \, s(t) \, G(\zeta,\Phi, \Psi)
\end{equation}
where $h^0$ is the amplitude of the incoming GW, $s(t)$ -- its normalized shape (e.g. sinusoidal), and $G(\zeta,\Phi,\Psi)$ -- the geometrical factor (G-factor) determined by
relative orientation of the antenna to the sky position of the GW source (angles $\zeta, \Phi$) at the fixed sidereal time (ST) of the event detection. $\Psi$ is the polarization angle of tensor wave (see definition \citealt{thorne87}).

\subsection{Tensor transverse GW}
In the general case of tensor transverse GW (spin-2 gravitons) the detected strain $\bar{h}(t; \zeta, \Phi, \Psi)$ contains combination of two polarizations: $h_+$ and $h_\times$, weighted by two antenna response functions $F_+$ and $F_\times$ in the proper reference frame:
\begin{equation} 
\label{eq:3}
    \bar{h} = F_+(\zeta, \Phi, \Psi) h_+(t) + F_\times(\zeta, \Phi, \Psi) h_\times(t)
\end{equation}
For a Michelson-type two-arms detector, the antenna-pattern functions $F_+, _\times$ are (\citealt{thorne87}, \citealt{will14}):
\begin{equation} \label{eq:5}
        F_+(\zeta, \Phi, \Psi) =\frac{1}{2} (1+\cos^2\zeta ) \cos 2\Phi \cos 2 \Psi - \cos \zeta \sin 2\Phi \sin 2\Psi
\end{equation}
\begin{equation} \label{eq:6}
        F_\times(\zeta, \Phi, \Psi) = \frac{1}{2} (1+\cos^2\zeta ) \cos 2\Phi \sin 2 \Psi + \cos \zeta \sin 2\Phi \cos 2\Psi
\end{equation}
Main designations are explained in Fig. (\ref{fig:sheme}). Antenna-patterns for tensor polarizations are illustrated in Figs. (\ref{fig:resp_twv1}) and (\ref{fig:resp_twv2}). 

According to (\ref{eq:1}), the G-factor for general case of tensor transverse GW at the fixed sidereal time $t$ is the composition of antenna-pattern functions $F_{+, \times}$ with the appropriate weight coefficients determined by the nature of the source. In the case of "pure" tensor polarization, whether "plus" or "cross", the G-factor is equal to the corresponding antenna-pattern function:
\begin{equation} \label{eq:11}
    \begin{split}
        &G_+(\zeta, \Phi, \Psi) = F_+(\zeta, \Phi, \Psi) \,, \\
        &G_\times(\zeta, \Phi, \Psi) = F_\times(\zeta, \Phi, \Psi)
    \end{split}
\end{equation}

\subsection{Scalar transverse and longitudinal GW}

\begin{figure}
\minipage{.7\linewidth}
    \centering
    \begin{subfigure}[b]{0.49\linewidth}
        \includegraphics[width=\textwidth]{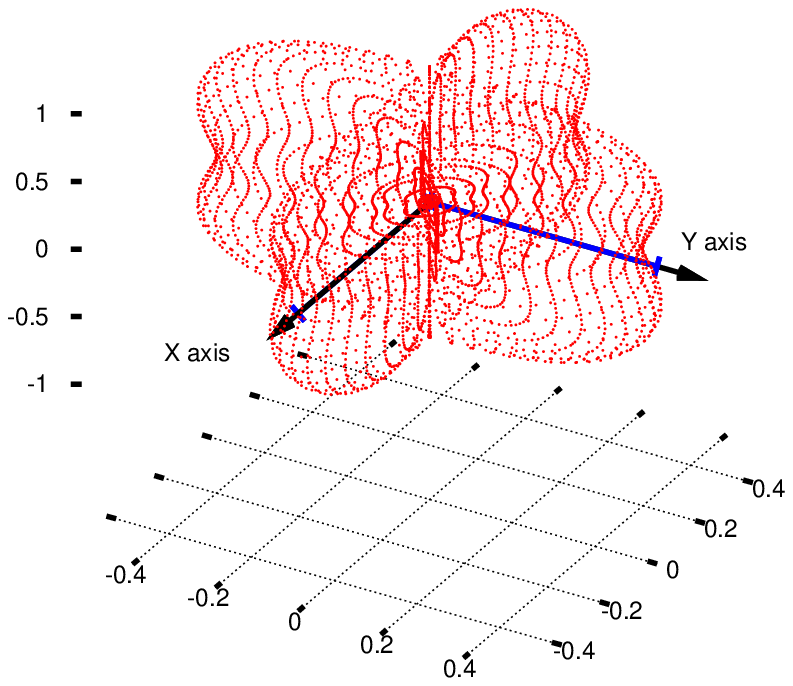}
        \caption{tensor "+" polarization}
        \label{fig:resp_twv1}
    \end{subfigure}
    \begin{subfigure}[b]{0.49\linewidth}
        \includegraphics[width=\textwidth]{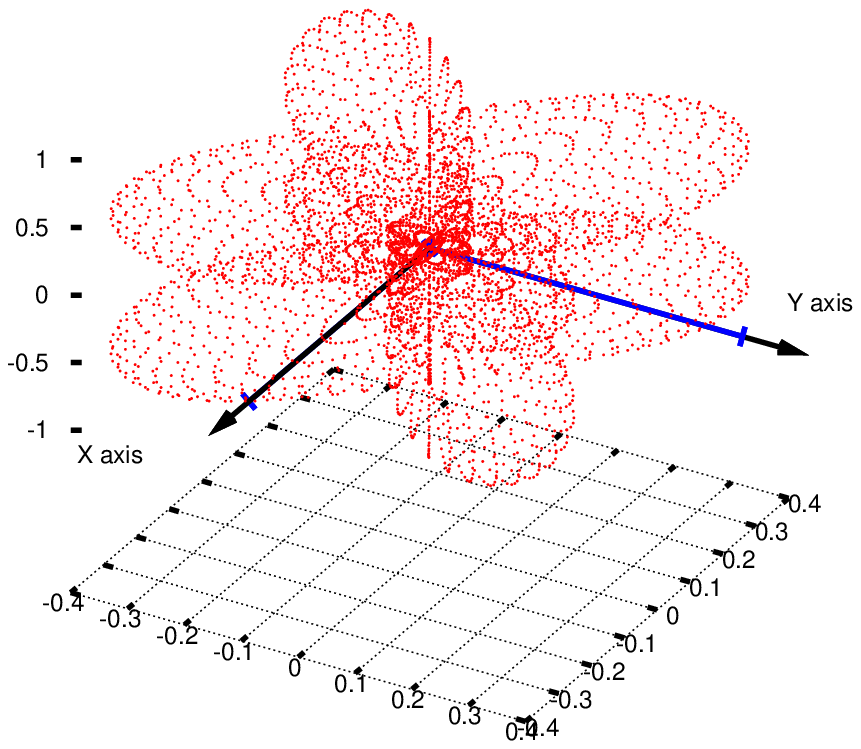}
        \caption{tensor "$\times$" polarization}
        \label{fig:resp_twv2}
    \end{subfigure}
    ~ 
    \begin{subfigure}[b]{0.49\linewidth}
        \includegraphics[width=\textwidth]{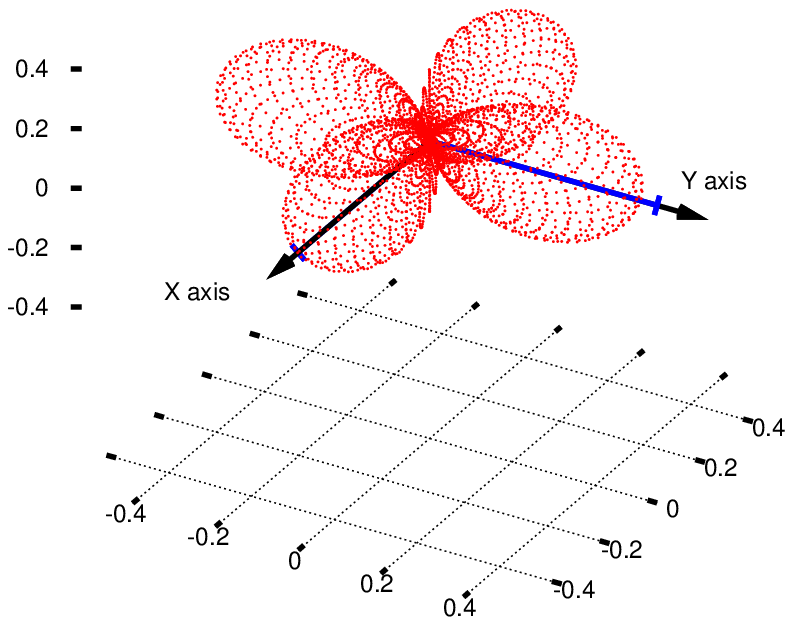}
        \caption{scalar waves for 2-arm antenna}
        \label{fig:resp_swv}
    \end{subfigure}
    \begin{subfigure}[b]{0.49\linewidth}
        \includegraphics[width=\textwidth]{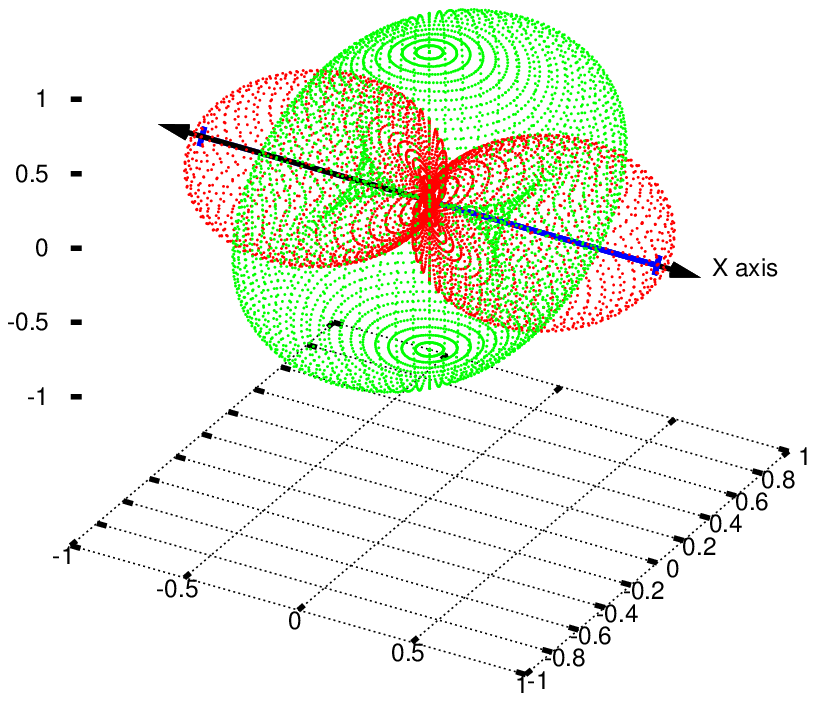}
        \caption{scalar transverse and longitudinal waves for 1-arm antenna}
        \label{fig:resp_swv_bar}
    \end{subfigure}
\endminipage
    \caption{Antenna patterns for different polarizations of an incoming GW. Blue lines indicate the arms of the detectors along the X and Y axis. Red points -- the beam pattern depending on the location of the GW source on the sky. Green points in (\ref{fig:resp_swv_bar}) -- the antenna response for scalar transverse wave in the case of one-arm mode. }\label{fig:resp}
\end{figure}

Geometrical factors for scalar longitudinal and transverse GW passing through
interferometric two-arms antenna are (\citealt{eardley73}, \citealt{will14}):
\begin{equation} \label{eq:9}
\begin{split}
    &G^{scal}_{long}(\zeta, \Phi) = \frac{1}{2} \sin^2\zeta \cos2\Phi\,, \\
    &G^{scal}_{trans}(\zeta, \Phi) = - \frac{1}{2} \sin^2\zeta \cos2\Phi
\end{split}
\end{equation}
Since G-factor of these states differs only by the sign, a two-arms antenna cannot disentangle between scalar transverse and longitudinal waves, Fig. (\ref{fig:resp_swv}).

\subsection{Application of one-arm antenna to the scalar polarization measurement}

To solve the problem of the indiscernibility between scalar longitudinal and transverse modes by a two-arms interferometer, we will consider the opportunity to use an antenna having one working arm with two test masses (one-arm mode). Then the observed strain is given by the length change of the working arm (X-axis) relative to the length $L_0$ of the fixed (former Y-axis) arm: $\Delta L(t) = L_X - L_0$. The amplitude of the arm-length variation $h^0 =\Delta L_\textrm{max}/L_0$ can be used as a normalization constant.

In the case of one-arm antenna, G-factors for scalar longitudinal and transverse GW (\citealt{baryshev01}) are:
\begin{equation}
\label{eq:12}
    G^{scal}_{long} = \cos\Theta = \sin\zeta \cos\Phi, \, \,  \, \, G^{scal}_{trans} = \sin\Theta
\end{equation}
where $\Theta$ -- the angle of the incidence, Fig. (\ref{fig:sheme}). The beam patterns are depicted on Fig. (\ref{fig:resp_swv_bar}), where the green points indicate response on a scalar transverse and red points -- on a scalar longitudinal GW. 
Consequently, it is possible to recognize scalar polarization modes by means of one-arm interferometric antenna.

\section{Results for the LIGO events} \label{two-antennas}

\subsection{Apparent circles of the allowed GW source positions}

\begin{figure}
\minipage{.8\linewidth}
  \includegraphics[width=\linewidth]{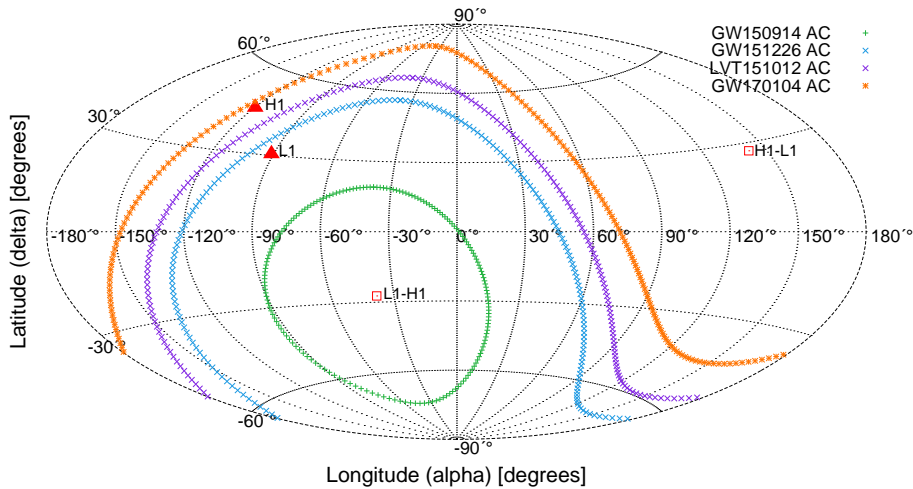}
\endminipage
  \caption{ACs of GW150914, LVT151012, GW151226 and GW170104 shown in Aitoff projection with respect to the Earth at the time of detection. Red triangles, H1 and L1, indicate the positions of the LIGO LI (Livingston) and H1 (Hanford) interferometers, H1-L1 and L1-H1 mark the poles of the line joining these two detectors (the points of the maximal time delay).}
    \label{fig:1eq_ai}
\end{figure}

\begin{table}
	\centering
	\caption{Detection parameters of LIGO events. ST is the sidereal time of the event, $h^0$ -- the strain as the maximal amplitude normalized by $10^{-21}$, $\Delta_\textrm{LH}$ -- the time delay between registrations at Livingston and Hanford antennas. ST is the sidereal time of the event given in hours.}
	\label{tab:params}
	\begin{tabular}{lccr} 
		\hline
		GW event (UTC) & ST & $\Delta_\textrm{LH}$ [ms] & $h^0$\\
		\hline
        GW150914    (09:50:45) & 3.33 & $6.9^{+0.5}_{-0.4}$  & 0.6  \\
        LVT151012   (09:54:43)   & 5.24 & $-0.6 \pm0.6$  & 0.3   \\
        GW151226    (03:38:53)   & 3.89 & $1.1 \pm0.3$  & 0.3  \\
        GW170104    (10:11:59)   & 11.1 & $-3.0^{+0.4}_{-0.5}$  & 0.3  \\
		\hline
	\end{tabular}
\end{table}

\begin{figure}
    \centering
    \begin{subfigure}[b]{0.8\linewidth}
        \includegraphics[width=\textwidth]{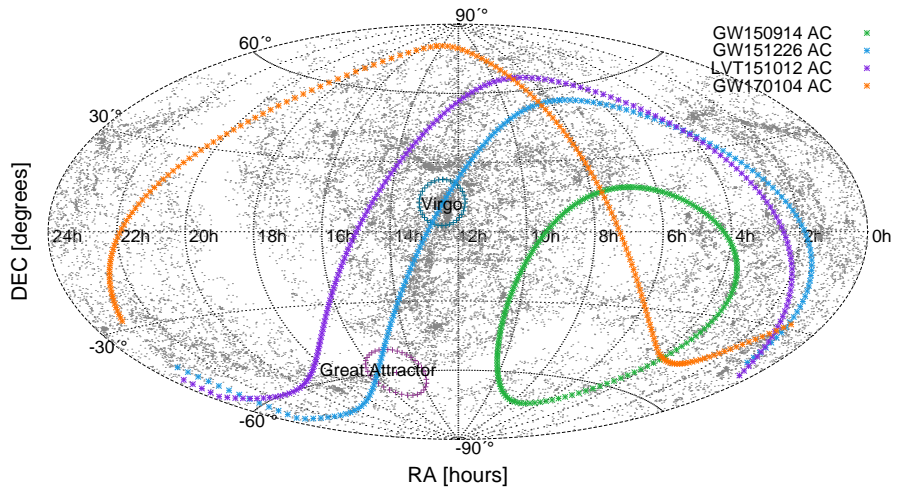}
        \caption{in equatorial coordinates}
        \label{fig:eq_ai}
    \end{subfigure}
    \begin{subfigure}[b]{0.8\linewidth}
        \includegraphics[width=\textwidth]{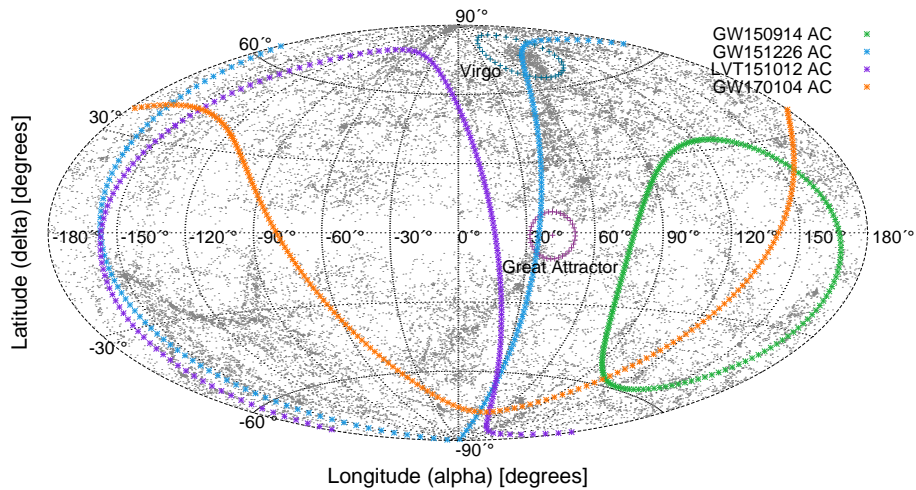}
        \caption{in galactic coordinates}
        \label{fig:gal_ai}
    \end{subfigure}
  \caption{ACs of the allowed source positions for the LIGO events: GW150914, LVT151012, GW151226 and GW170104 in equatorial and galactic coordinate systems (Aitoff projection).}
\end{figure}

Let us consider a certain LIGO event detected by two LIGO antennas: Livingston (L1) and Hanford (H1). The detected time delay $\Delta_\textrm{LH}$ between registrations at these two antennas determines a radius of an apparent circle (hereafter AC) on the unit sky sphere, along which possible sources of GW might be located. The centre of the AC is defined by the direction of the vector L1 -- H1 joining the two antennas at the sidereal time (ST) of the event. In Fig. (\ref{fig:1eq_ai}) we present the ACs for detected by LIGO events GW150914, LVT151012, GW151226 and GW170104 in the first equatorial coordinate system with respect to the Earth, as it is also shown by the LIGO-Virgo Scientific Collaborations (\citealt{abbott16a}).

\subsection{The specific role of the Local Super-Cluster plane}

We have considered three coordinate systems: equatorial, galactic and supergalactic (hereafter SG). Here we present our results of AC calculations in the supergalactic coordinate system (SG), which
has the North Pole $SGB = 90^{\circ}$ with galactic coordinates
$l = 47.37^{\circ}$, $b = 6.32^{\circ}$ (\citealt{courtois13}). 

In Figs. (\ref{fig:eq_ai}, \ref{fig:gal_ai}) the ACs of the allowed GW sources are shown with the background projection
of the 2MRS catalog of galaxies (\citealt{huchra12}), which is the result of the 2MASS all-sky IR survey and includes the redshifts of $43 \, 533$ galaxies. 
We have taken the subsample containing $32 \, 656$ galaxies with $z \leq 0.025$. Our sample corresponds to the spatial distribution of the galaxies within $\sim 100$ Mpc known as the Local Super-Cluster or Laniakea Super-Cluster or Home Super-Cluster (hereafter LSC, \citealt{devaucou58},
\citealt{paturel88},
\citealt{dinella94},
\citealt{courtois13},
\citealt{tully14}).
The LSC has the filamentary disc-like structure with the radius $\sim 100$ Mpc, the thickness $\sim 30$ Mpc and the centre roughly in the Virgo cluster ($SGL = 104^{\circ}; SGB = 22^{\circ}$).

Interestingly, the apparent circles for three GW events detected in 2015 lie along the supergalactic plane of the Local Super-Cluster of galaxies (Fig.(\ref{fig:twv_sg_ai})), which may indicate a special
role of the LSC regarding to these events. Whereas the AC for GW170104 lies perpendicularly to the SG plane, it has some parts within LSC, thus allowing to take into consideration the possibility that its GW source may also belong to the LSC. 
This supposition is consistent with the detection of the GRB-like afterglow ATLAS17aeu (\cite{stalder17}, indicated by the green triangle in all ACs figures in SG coordinates), which might be identified as a possible source of GW170104.

\subsection{G-factors along an apparent circle}

For each possible place of the GW source along the considered AC the G-factor $G(\Phi, \zeta, \Psi)$ can be calculated using general formulas (\ref{eq:5}, \ref{eq:6}) for tensor GW or (\ref{eq:9}, \ref{eq:12}) -- for scalar GW. The $G$-factor depends upon the azimuth $\Phi$ and zenith angle $\zeta$ of the considered point in the horizontal coordinate system of an antenna (see Fig. \ref{fig:sheme}) as well as on the polarization angle $\Psi$ in the case of tensor wave. Therefore, for each point $G_L \equiv G(\Phi_L, \zeta_L, \Psi_L)$ and $G_H \equiv G(\Phi_H, \zeta_H, \Psi_H)$ should be calculated for Livingston $L1$ and Hanford $H1$ antennas separately.
Then the expected value of the strain ratio:
\begin{equation} \label{eq:14}
    \frac{h_L}{h_H} = \frac{G_L}{G_H} = \frac{G(\Phi_L, \zeta_L, \Psi_L)}{G(\Phi_H, \zeta_H, \Psi_H)}
\end{equation}

\begin{figure}
\centering
    \begin{subfigure}[b]{.6\linewidth}
        \includegraphics[width=\linewidth]{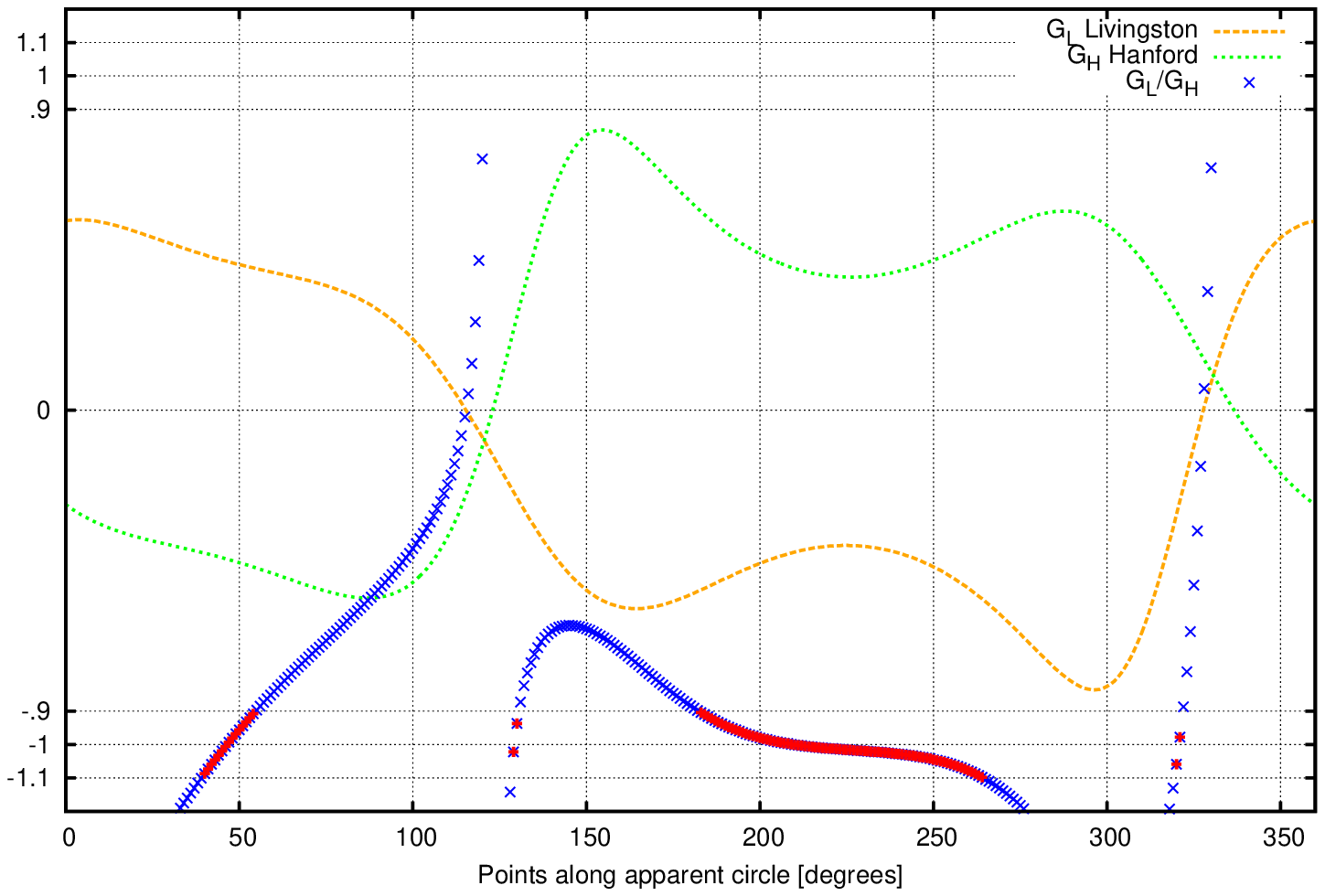}
        \caption{for tensor "+" GW}
        \label{fig:g_twv_GW150914}
    \end{subfigure}
    \begin{subfigure}[b]{.6\linewidth}
        \includegraphics[width=\linewidth]{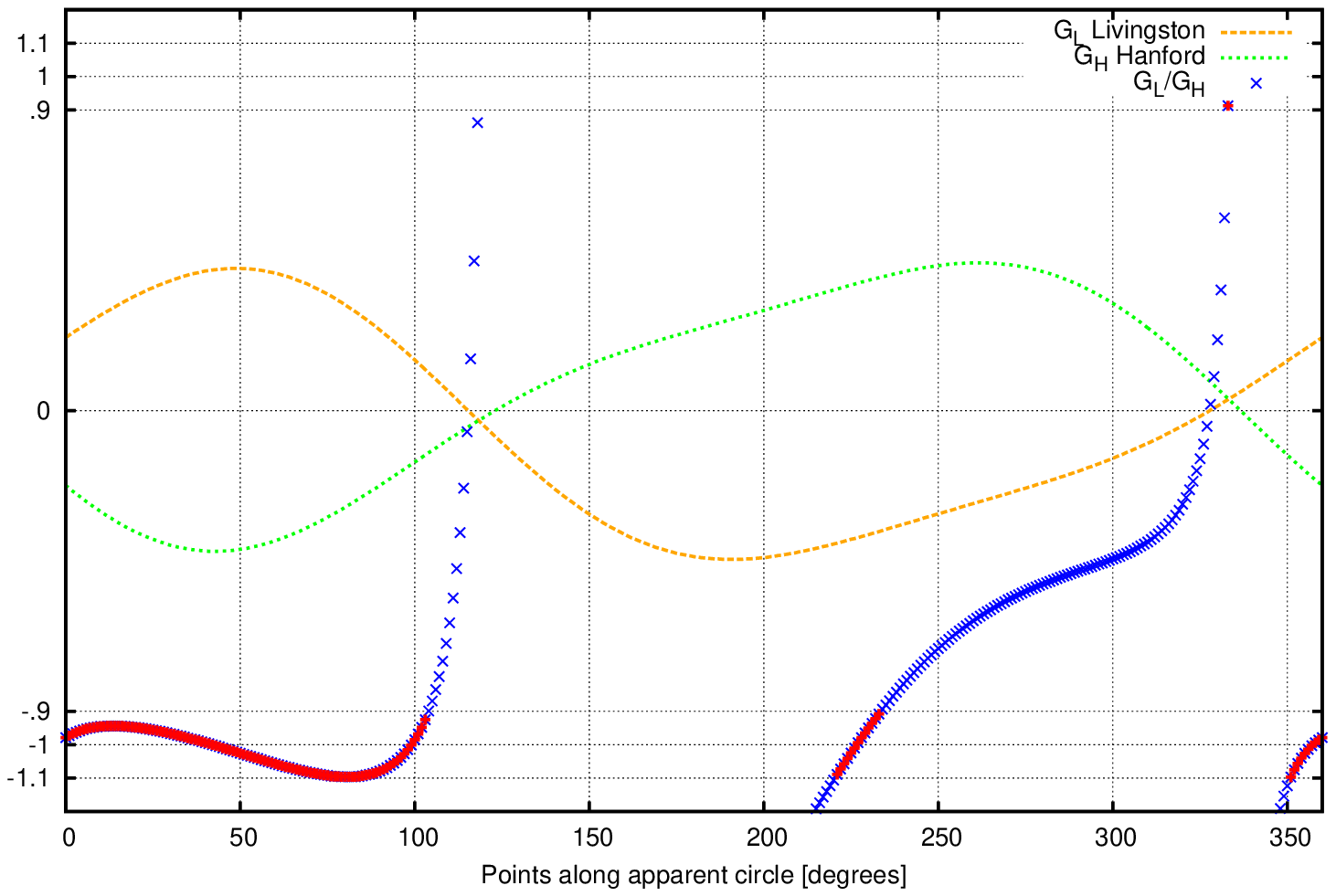}
        \caption{for scalar (transverse or longitudinal) GW}
        \label{fig:g_swv_GW150914}
    \end{subfigure}
    \begin{subfigure}[b]{.6\linewidth}
        \includegraphics[width=\linewidth]{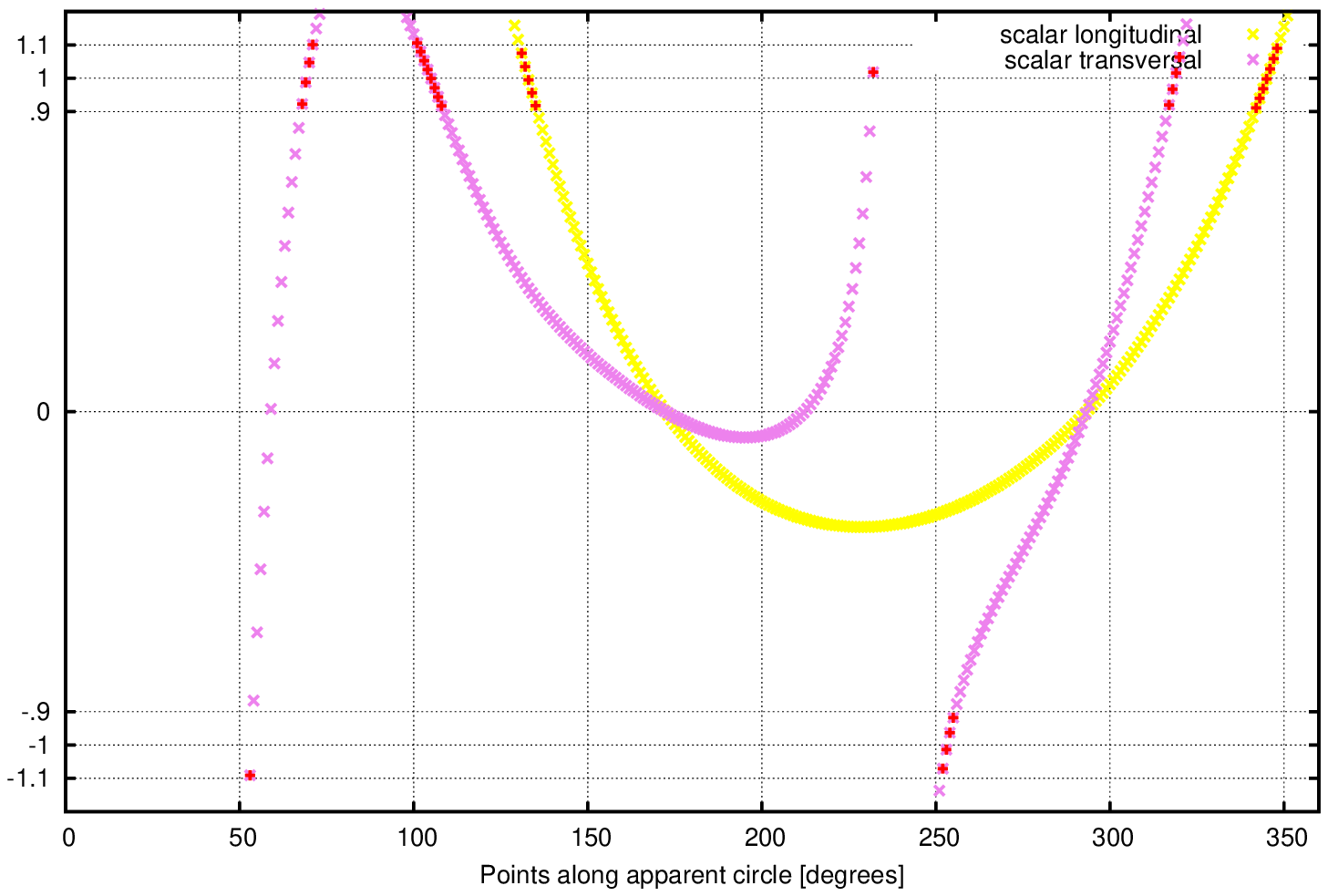}
        \caption{for scalar GW as it would be detected in one-arm mode. The yellow curve represents the ratio $G_L/G_H$ for scalar longitudinal and the violet -- for scalar transverse wave.}
        \label{fig:g_swv_bar_GW150914}
    \end{subfigure}
        \caption{G-factors for possible polarization states of the GW150914. The orange and green thin curves represent the G-factors along the AC calculated with LIGO Livingston and Hanford coordinates respectively. Blue dots curve shows the ratio $G_L/G_H$. The allowable places of the GW source are highlighted by red corresponding to the condition $G_L/G_H \approx 1 \pm 10\%$.
        Figs. (\ref{fig:g_twv_GW150914}, \ref{fig:g_swv_GW150914}) use the data by currently working two-arms interferometric anntennas LIGO, while (\ref{fig:g_swv_bar_GW150914}) illustrates the possible results in the case of one-arm detectors (see the explanation in the text). }
\end{figure}

%
%
Thus, the calculated ratio of $G_L/G_H$ for a certain point on an AC predicts the observed strain ratio $h_L/h_H$. Therefore, it is possible to highlight such points on the AC, where the calculated $G_L/G_H$ approximately equal to the observed $h_L/h_H$. 

%

$G$-factors for all detected events were calculated for the considered above cases of tensor and scalar waves. The Figs. (\ref{fig:g_twv_GW150914}, \ref{fig:g_swv_GW150914}, \ref{fig:g_swv_bar_GW150914}) illustrate calculation and selection process for the GW150914 event. The green and yellow curves indicate $G$-factors of the considered GW polarization calculated for each point along the AC with respect to its horizontal coordinates at Livingston and Hanford antennas. Blue dots curve shows the ratio $G_L/G_H$, where red points indicate the most probable places according to the condition $G_L/G_H \approx 1 \pm 10\%$.

In the case of tensor transverse waves, calculations were made using formulae (\ref{eq:11}) for "+" polarization with $\Psi = 0^{\circ}$. For scalar wave detected by two-arms interferometric antennas, the equations (\ref{eq:9}) give the same results for both longitudinal and transverse mode. As it has been shown by formulas (\ref{eq:12}) and antenna response graphics Fig. (\ref{fig:resp}), only one-arm antennas allow us to distinguish between scalar longitudinal and transverse GW. In this antenna mode, Fig. (\ref{fig:g_swv_bar_GW150914}), yellow dots curve shows the $G_L/G_H$ for scalar longitudinal, and violet -- for scalar transverse wave. As in other cases, red points indicate most probable places of GW sources along the AC.

\begin{figure}
  \includegraphics[width=\linewidth]{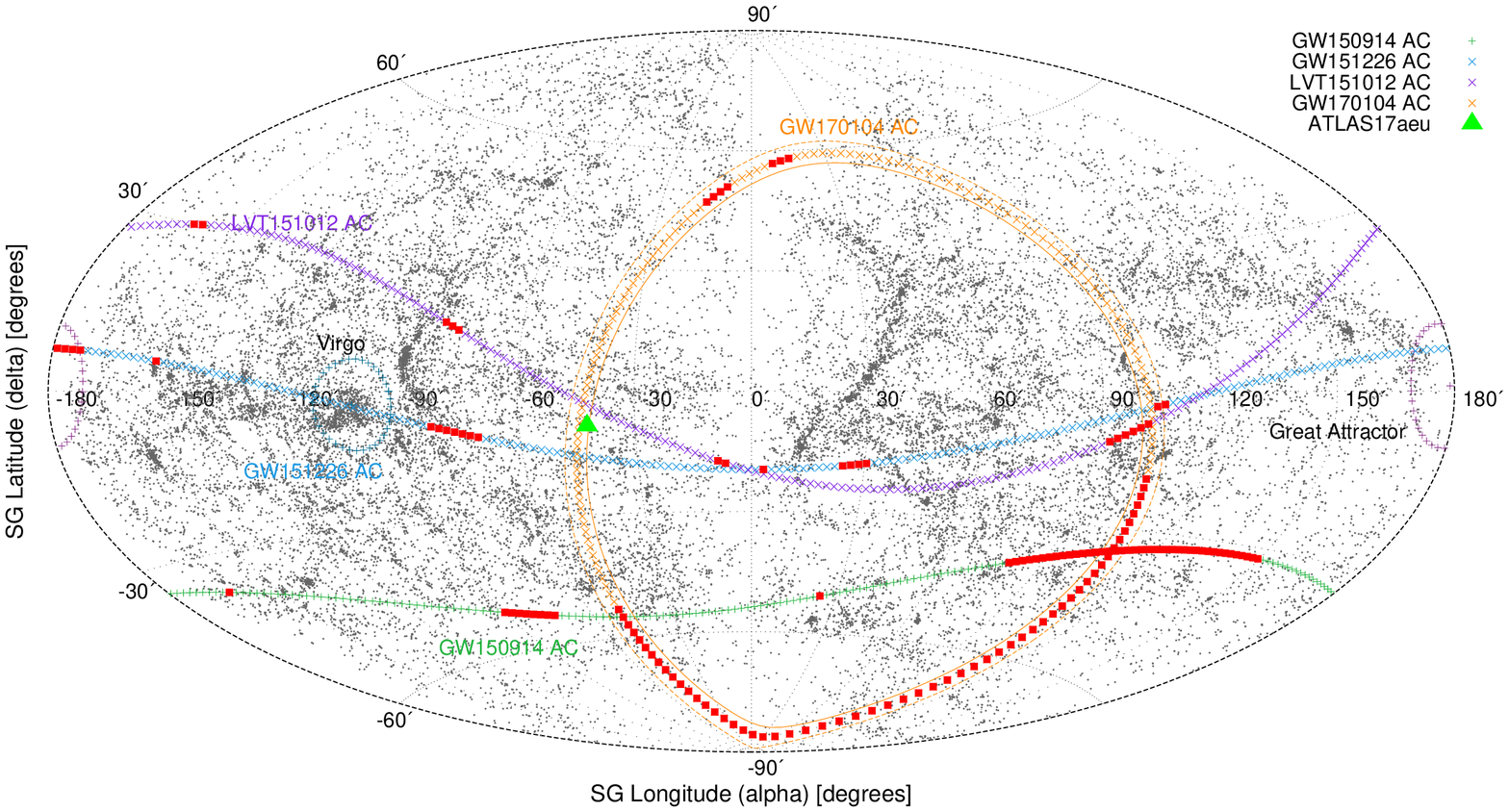}
  \caption{ACs for LIGO events in supergalactic coordinates. The red points corresponding to the condition $G_L/G_H \approx 1 \pm 10\%$ represent the allowed source positions in the case of tensor "+" ($\Psi=0$) incoming GW. The green triangle denotes the possible optical counterpart ATLAS17aeu for the GW170104.}
    \label{fig:twv_sg_ai}
\hfill
  \includegraphics[width=\linewidth]{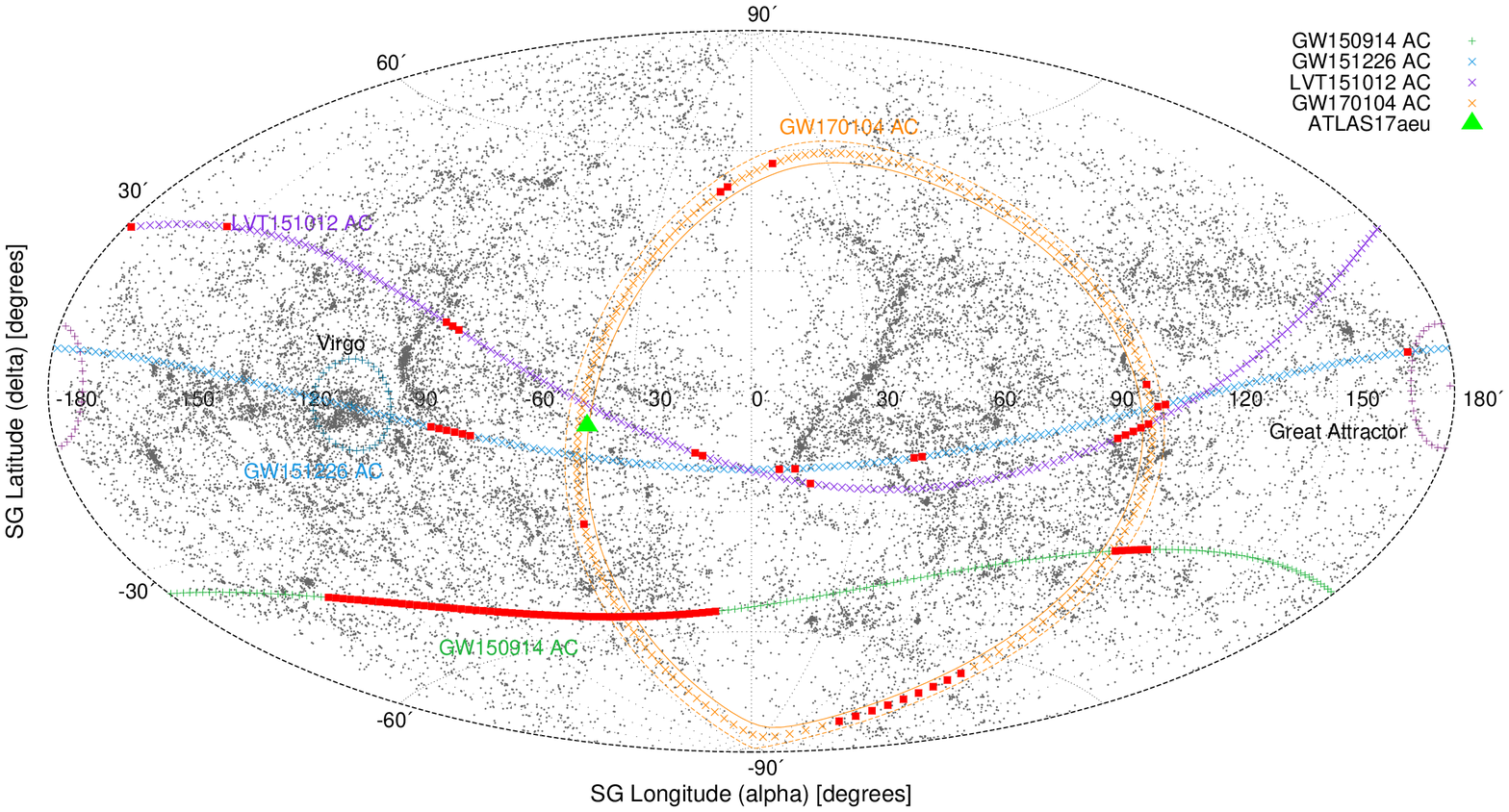}
  \caption{ACs for LIGO events in supergalactic coordinates. The red points corresponding to the condition $G_L/G_H \approx 1 \pm 10\%$ represent the allowed source positions in the case of scalar (transverse or longitudinal) incoming GW. The green triangle denotes the possible optical counterpart ATLAS17aeu for the GW170104.}
    \label{fig:swv_sg_ai}
\end{figure}


\begin{figure}
  \includegraphics[width=\linewidth]{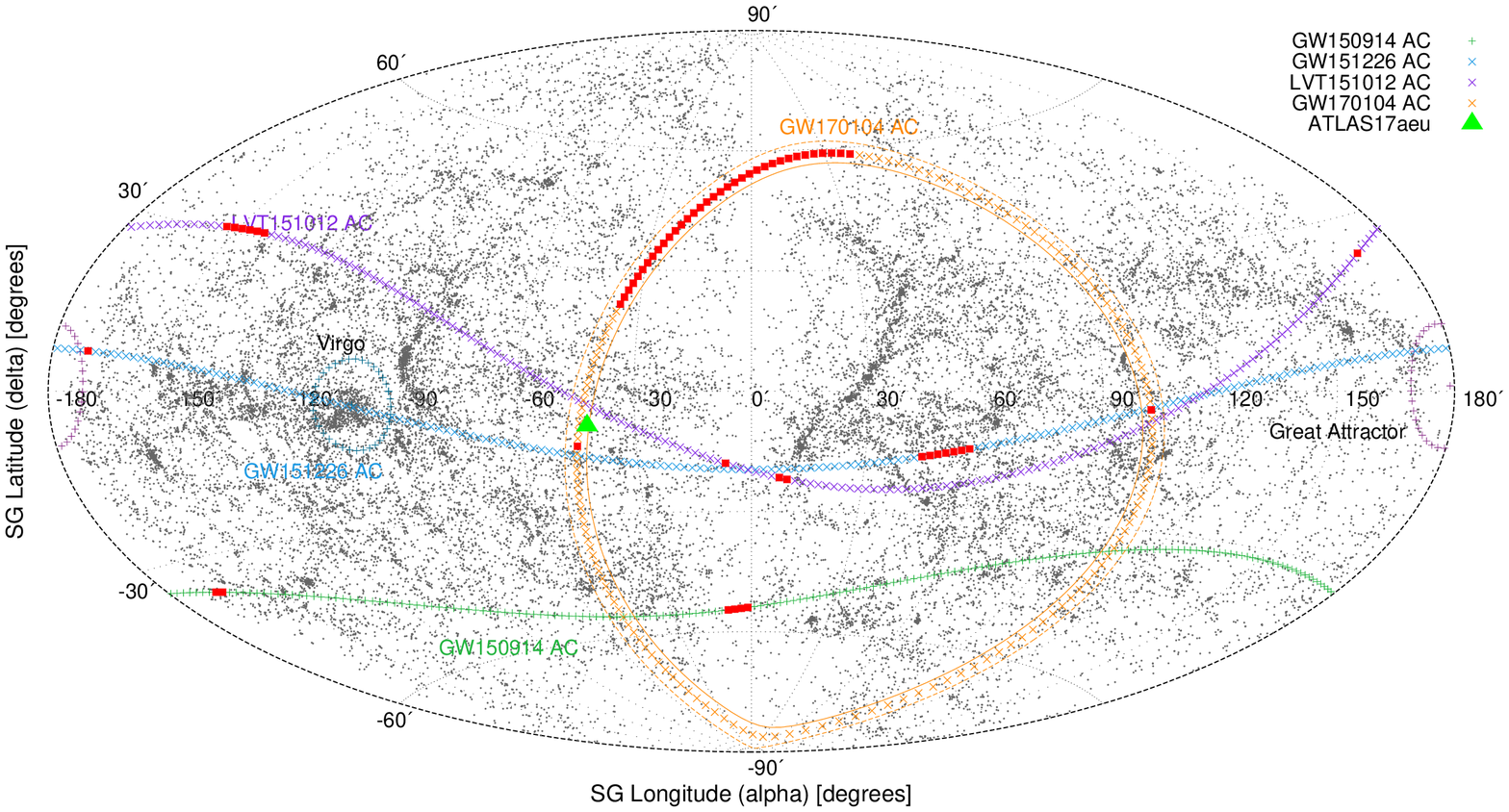}
  \caption{ACs of the predicted source positions for LIGO events in the case of scalar longitudinal GW and 1-arm interferometric detector in SG coordinates. Red points correspond to the condition $G_L/G_H \approx 1 \pm 10\%$. The green triangle denotes the possible optical counterpart ATLAS17aeu for the GW170104.}
    \label{fig:slw_sg_ai}
\hfill
  \includegraphics[width=\linewidth]{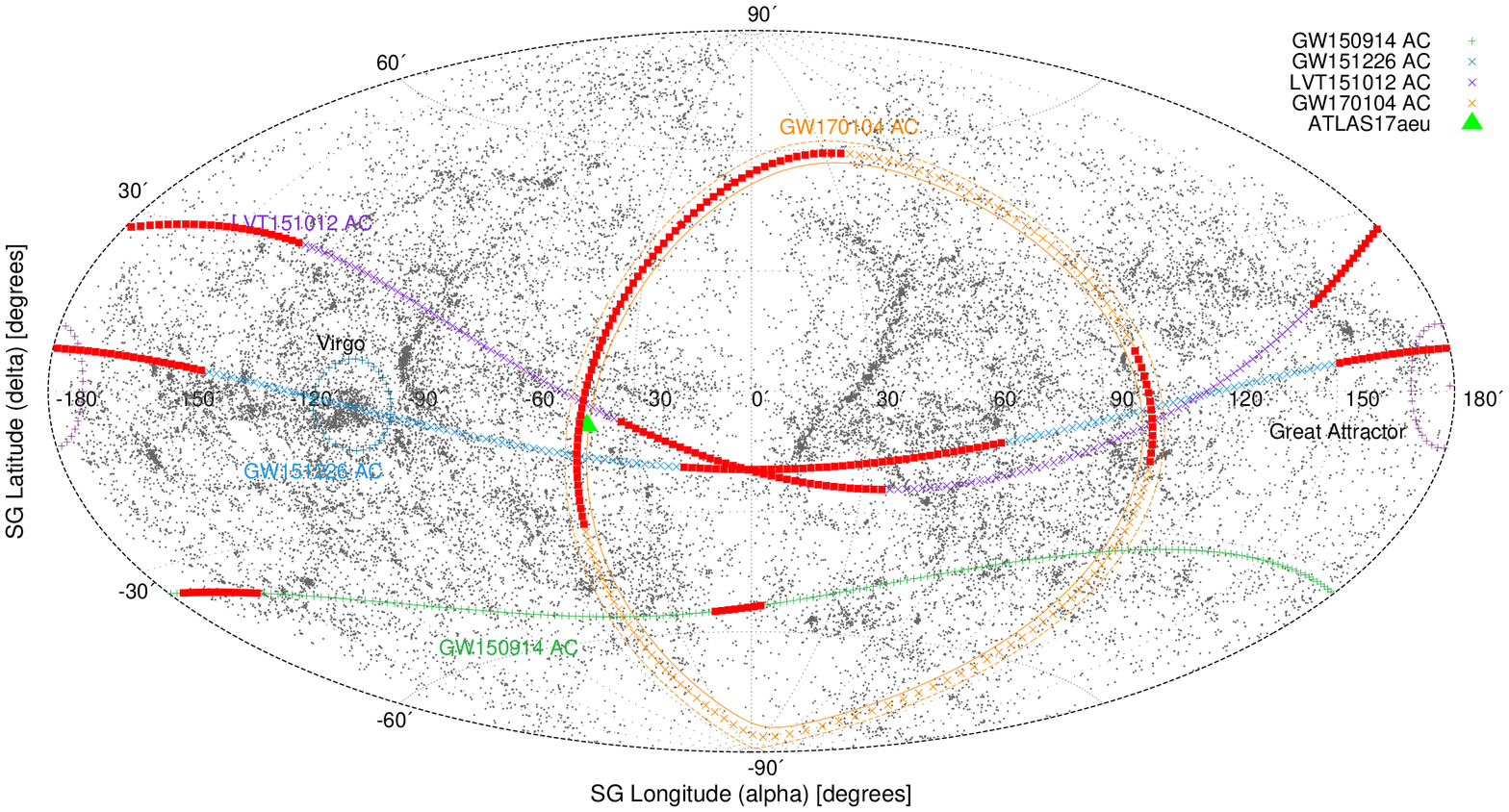}
  \caption{ACs of the predicted source positions for LIGO events in the case of scalar transverse GW and 1-arm interferometric detector in SG coordinates. Red points correspond to the condition $G_L/G_H \approx 1 \pm 10\%$. The green triangle denotes the possible optical counterpart ATLAS17aeu for the GW170104.}
    \label{fig:stw_sg_ai}
\end{figure}

\begin{figure}
  \includegraphics[width=\linewidth]{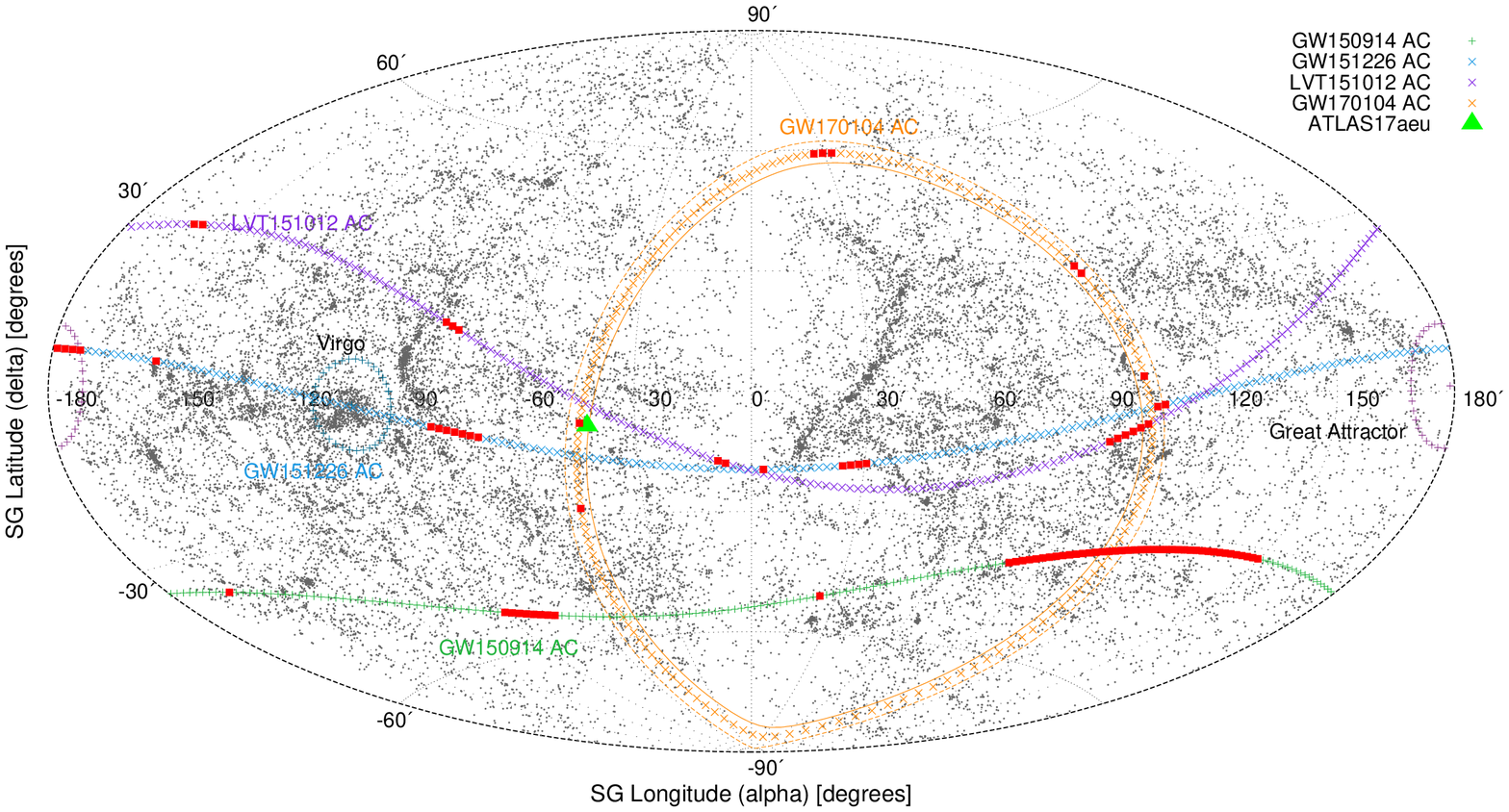}
  \caption{ACs of the predicted source positions for LIGO events in the case of tensor transverse GW ($G = 1.5 F_+  + \sqrt{2} F_\times$) in SG coordinates. Red points correspond to the condition $G_L/G_H \approx 1 \pm 10\%$. The green triangle denotes the possible optical counterpart ATLAS17aeu for the GW170104.}
    \label{fig:twv_mix_sg_ai}
\end{figure}

Finally, the ACs of GW150914, LVT151012, GW151226 and GW170104 with the predicted sky localization according to the method (highlighted by red) are shown on Fig. (\ref{fig:twv_sg_ai}) for tensor ''+'' and Fig. (\ref{fig:swv_sg_ai}) -- for scalar incoming GW. Additionally, Figs. (\ref{fig:slw_sg_ai}) and (\ref{fig:stw_sg_ai}) depict the allowed places of sources in the case of scalar longitudinal and transverse waves respectively, as it would be detected by an one-arm antenna.

The detected GRB-like afterglow ATLAS17aeu (\cite{stalder17}) as a possible counterpart event of the GW170104 is indicated by the green triangle. The AC of the GW170104 is depicted taking into account the error in the determination of the time delay $\pm 0.5$ ms, Tab. (\ref{tab:params}), which allows us to represent the realistic possible area for the forthcoming search of electromagnetic transients for this event. Besides this, Fig. (\ref{fig:twv_mix_sg_ai}) presents the case of a mixture of tensor transverse modes with $G = 1.5 F_+  + \sqrt{2} F_\times$, where the covering area is coincide with the position of the ATLAS17aeu. Thereby focusing on the fact that there may exist such polarization states of an incoming GW providing various localization areas including those within the SG plane.


\section{Application of the three antennas LIGO-Virgo to the GW polarisation state determination}

\begin{figure}
    \centering
    \begin{subfigure}[b]{\linewidth}
        \includegraphics[width=\linewidth]{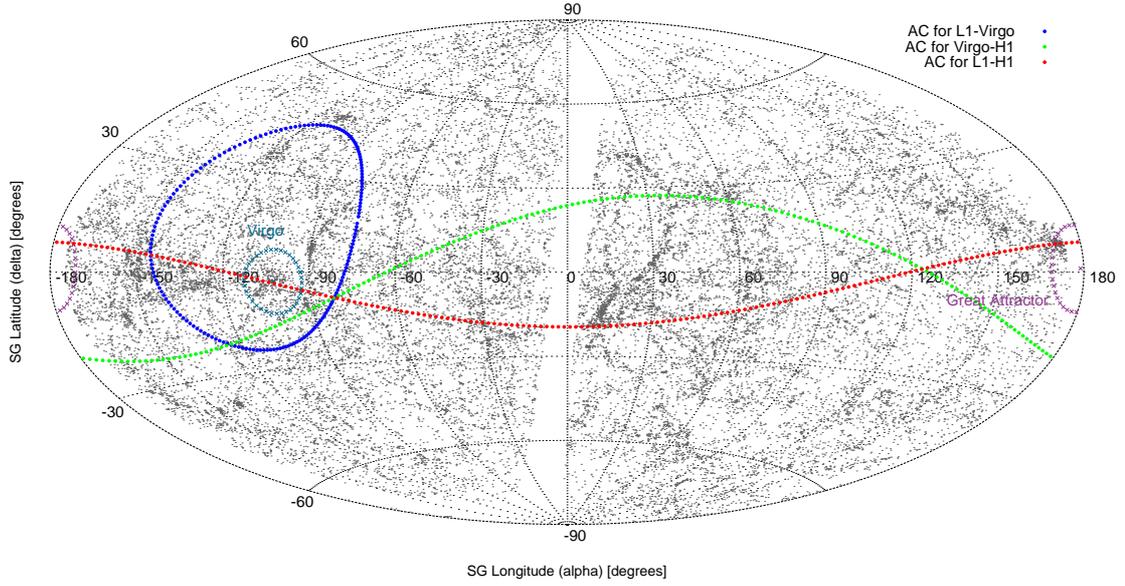}
        \caption{The source coordinates RA $= 11.6^h;$ DEC $=30.3^\circ$. }
        \label{fig:1src_sg_ai}
    \end{subfigure}
    \begin{subfigure}[b]{\linewidth}
        \includegraphics[width=\linewidth]{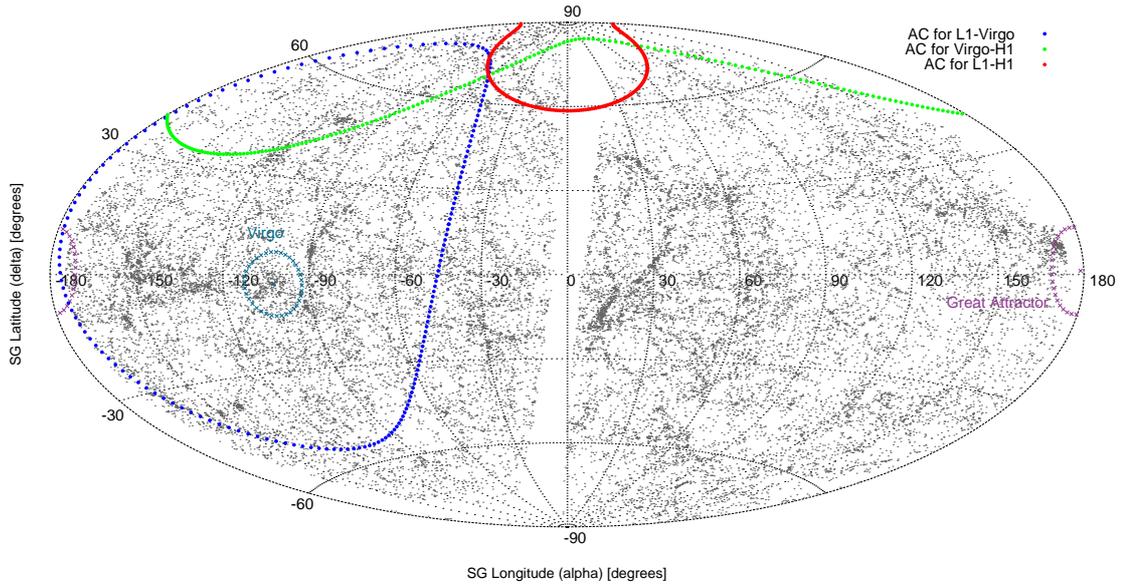}
        \caption{The source coordinates RA $= 18^h$, DEC $=32^\circ$. }
        \label{fig:2src_sg_ai}
    \end{subfigure}
    \caption{The apparent circles intersecting in the point of an artificial GW source in the case of detection by the three antennas: LIGO Livingston, LIGO Hanford and Virgo (L-H-V), at the time of the event GW151226. }
\end{figure}

For the reported GW events (\citealt{abbott16a}), there were only two antennas, LIGO Hanford and Livingston, in operation. In such a case, the localization of the source may be considered not in a point, but rather along an apparent circle of the event. Otherwise, when three antennas are operating, such as LIGO L1, H1 (USA) and Virgo (Italy), it is possible to localize the GW source in a point of three ACs intersection (Fig.\ref{fig:1src_sg_ai}). Thereby, the information about the ratio of amplitudes on each pair of antennas allows us to distinguish between different types of GW polarization: scalar and tensor, transverse and longitudinal. The details about localization of the antennas on the Earth are given in Tab. (\ref{tab:3det}).

\begin{table}
	\centering
	\caption{Parameters of the LIGO and Virgo GW antennas}
	\label{tab:3det}
	\begin{tabular}{lccr} 
		\hline
		Name & Latitude & Longitude; Azimuth \\
        LIGO L1 & $46^{\circ}{27}'{19}''$N & $119^{\circ}{24}'{28}''$W; N$36^{\circ}$W \\
        LIGO H1 & $30^{\circ}{33}'{46}''$N & $90^{\circ}{46}'{27}''$W; W$18^{\circ}$S  \\
        Virgo & $43^{\circ}{37}'{53}''$N & $10^{\circ}{30}'{16}''$E; N$19^{\circ}$E \\
		\hline
	\end{tabular}
\end{table}
\begin{table}
	\centering
	\caption{Calculated G-factors for two artificial GW-sources, which could be detected by LIGO Livingston, LIGO Hanford and Virgo (L-V-H) at the time
  of GW151226 event. The labels such as $G_\textrm{1/2}$ denote the ratio $G_1/G_2$ with respect to a couple of antennas. The time delays $\Delta$ are given in {[}ms{]}.}
	\label{tab:3amp}
	\begin{tabular}{lccccccccr} 
		\hline
		Polarization state & $G_\textrm{L}$ & $G_\textrm{H}$ & $G_\textrm{V}$ & $G_\textrm{L/H}$ & $G_\textrm{L/V}$ & $G_\textrm{V/H}$ & $\Delta_\textrm{LH}$ & $\Delta_\textrm{LV}$ & $\Delta_\textrm{VH}$   \\
		\hline
        \multicolumn{7}{l}{RA $= 11.6^h$ DEC $=30.3^\circ$} & 1.09 & -23.35 & -0.97 \\
		\hline
        Tensor +, $\Psi = 0$ & 0.376 & -0.356 & -0.608 & -1.06 & -0.62 & 1.71 \\
        Scal. long. or trans. & 0.373 & -0.346 & -0.029 & -1.08 & -12.86 & 0.08 \\
        Scal. long. (1-arm) & -0.933 & 0.384 & -0.128 & -2.43 & 7.29 & -0.33 \\
        Scal. trans. (1-arm) & 0.359 &  0.924 & 0.992 & 0.39 & 0.36 & 1.07 \\
		\hline
        \multicolumn{7}{l}{RA $= 18^h$ DEC $=32^\circ$}   &  -9.49 & -11.53 & -25.23  \\
		\hline
        Tensor +, $\Psi = 0$ & -0.525 & 0.468 & 0.190 & -1.12 & -2.76 & 0.41 \\
        Scal. long. or trans. & -0.409 & 0.460 & 0.181 & -0.89 & -2.26 & 0.39 \\
        Scal. long. (1-arm) & 0.169 & 0.977 & 0.819 & 0.17 & 0.21 & 0.84 \\
        Scal. trans. (1-arm) & 0.985 & 0.212 & 0.574 & 4.65 & 1.72 & 2.71 \\
		\hline
	\end{tabular}
\end{table}

To demonstrate a possibility for distinction between different GW polarizations by means of three LIGO-Virgo antennas, let us consider two artificial sources: one located near the SG plane and another one -- outside the SG plane. As a concrete test, we have taken these artificial sources with equatorial coordinates RA $= 11.6^h$, DEC $=30.3^\circ$, near the Virgo cluster, and with RA $= 18^h$, DEC $=32^\circ$ (SG $B = 70^\circ$), i.e. outside the SG plane. The time of the artificial event has been proposed to be the same as the time of the GW151226.

The ACs for these two artificial sources are shown in supergalactic coordinates on the Figs. (\ref{fig:1src_sg_ai}, \ref{fig:2src_sg_ai}). Red curves indicate the ACs constructed with respect to the couple LIGO Livingston -- Hanford (L1-H1), blue -- L1-Virgo, and green -- Virgo-H1.

The G-factors for these artificial sources were calculated for each detector in the network. According to the condition (\ref{eq:14}), the ratio of the G-factors at each couple of the antennas represents the strain ratio to be detected at the time of GW151226, Tab. (\ref{tab:3amp}). Additionally, predictions for strain ratio at one-arm antenna clearly illustrate the possibility of recognition scalar polarization mode by means of such antenna construction.

To sum up, in the case when the localization of a GW source is known by means of three and more detectors in operation, it is possible to make assumptions about polarization state of the wave using the information about the detected strain ratio and matching it with the calculated ratio of G-factors.

\section{Conclusions}

We have presented the new method for a GW source localization on the sky in the case of a GW detection by two interferometric antennas. The method is based on the antennas beam patterns for a supposed polarization state of the incoming GW together with the measurements of the arrival time delays between antennas and the ratio of the detected strains at each antenna.


It has been shown that a network of LIGO-type two-arms antennas can distinguish between tensor and scalar, but not between scalar longitudinal and transverse polarizations, which is possible by means of one-arm interferometric (as well as bar) detectors.

We have demonstrated that there is an interesting possibility for the polarization state recognition by means of actual localization on the sky a GW source. For this purpose, there has been considered a network of three antennas LIGO-Virgo. The theoretical conclusions concerning beam patterns for different polarization states were applied to the calculations of the strain ratio for each antenna couple in order to offer a test on the polarization state of the GW coming from a definitely located source.

For three aLIGO events: GW150914, GW 151226 and LVT151012,
the apparent circles of the allowed GW source positions are parallel to the supergalactic (SG) plane of the Local Super-Cluster of galaxies. Such fact indicates that GW sources of these events might belong to this structure. It is worth noting that if the detected three events did not belong the LSC, then we would have a rare chance of accidentally correlated direction of GW sources positions on the sky, especially for the sources at very high distances such the currently proposed $400 \div 1000$ Mpc for the LIGO events 2015 (\citealt{abbott16a}, \citealt{abbott16b}). Moreover, if these GW sources are related to the LSC, then we have to consider distances to them within $\sim 100$ Mpc.

Interestingly, for the new aLIGO event GW 170104, the apparent circle is perpendicular to the supergalactic equator with only some parts within SG plane ($\pm 30^\circ$ SGB), nevertheless the possible optical counterpart ATLAS17aeu to this event (\citealt{stalder17}) belongs to the Local Super-Cluster plane, which is also consistent with our supposition about the special role of the LSC.

The next aLIGO observing runs are proposed to test the reality of clustering the GW events along the SG plane. Future identification of GW sources with electromagnetic counterparts is crucial for the physics of the gravitational interaction.  Especially, follow-up observations such as \citealt{copper16}, \citealt{stalder17} are of great importance for the fundamental physics and they
should take into account the experience in GRB optical identification (such as \citealt{vlasyuk2016},  \citealt{castro16}).

\section*{Acknowledgements}
We thank referee for the useful comments, which help us to improve the presentation of our results. This work was supported by the Saint Petersburg
State University.

\noindent

\nocite{*}
\bibliographystyle{amsplain}

\end{document}